\documentclass{PoS}
 
\usepackage{amssymb,amsmath}
\usepackage{graphicx}

\title{RadioLensfit: Bayesian weak lensing measurement in the visibility domain}

\ShortTitle{RadioLensfit}

\author{\speaker{Marzia Rivi}$^{a,b}$, Lance Miller$^{b}$, Sphesihle Makhathini$^{c,d}$, Filipe Batoni Abdalla$^{a,c}$\\
        \llap{$^{a}$}Department of Physics and Astronomy, University College London, Gower Street, London, UK\\
        \llap{$^{b}$}Astrophysics, Department of Physics, University of Oxford, Keble Road, Oxford, UK\\
        \llap{$^{c}$}Department of Physics and Electronics, Rhodes University, Grahamstown, 6140, South Africa \\
        \llap{$^{d}$}SKA South Africa\\
        E-mail: \email{m.rivi@ucl.ac.uk}}

\abstract{Observationally, weak lensing has been served so far by optical surveys due to the much larger number densities of background galaxies achieved, which is typically by two to three orders of magnitude larger compared to radio. However, the high sensitivity of the new generation of radio telescopes such as the Square Kilometre Array\footnote{https://www.skatelescope.org/} (SKA) will provide a density of detected galaxies that is comparable to that found at optical wavelengths, and with significant source shape measurements to make large area radio surveys competitive for weak lensing studies.
This will lead weak lensing to become one of the primary science drivers in radio surveys too, with the advantage that they will access the largest scales in the Universe going beyond optical surveys, like LSST\footnote{http://www.lsst.org/} and Euclid\footnote{http://www.euclid-ec.org/}, in terms of redshifts that are probed. 
RadioLensfit is an adaptation to radio data of \emph{lensfit}, a model-fitting approach for galaxy shear measurement, originally developed for optical weak lensing surveys. Its key advantage is working directly in the visibility domain, which is the natural approach to adopt with radio data, avoiding systematics due to the imaging process.
We present results on galaxy shear measurements, including investigation of sensitivity to instrumental parameters such as the visibilities gridding size, based on simulations of individual galaxy visibilities performed by using SKA1-MID baseline configuration. We get an amplitude of the shear bias in the method comparable with SKA1 requirements for a population of galaxies with realistic flux and scalelength distributions estimated from the VLA SWIRE catalog.}

\FullConference{EXTRA-RADSUR2015 (*)\\
		20--23 October 2015\\
		Bologna, Italy

                \bigskip
                \hrule
                \bigskip

                \textnormal{(*) This conference has been organized
                  with the support of the Ministry of Foreign Affairs
                  and International Cooperation, Directorate General
                  for the Country Promotion (Bilateral Grant Agreement
                  ZA14GR02 - Mapping the Universe on the Pathway to
                  SKA)}
}

\begin{document}
\section{Introduction}
\label{sec:intro}

Currently, techniques available for the measurement of cosmic shear (the galaxy shape distortion due to a gravitational potential on large scales) are based on the best fitting of galaxy images, as they have been developed for optical surveys. 
However the image reconstruction from radio observations via PSF deconvolution, using algorithms such as CLEAN, is a highly non-linear process and has the potential to produce spurious cosmic shear. Moreover the noise in radio images is highly correlated. Therefore, accurate methods for radio weak lensing should work in the visibility domain, where the noise is Gaussian and there is a perfect modelling of the sampling function.

At present the only investigations about shear measurement with radio sources use shapelets, where galaxy shape models are decomposed through an orthonormal basis of functions corresponding to perturbations around a circular Gaussian (for an overview see~\cite{Patel15}). 
In this paper we propose to use a more realistic galaxy model and to adapt a Bayesian model fitting approach called \emph{lensfit}~\cite{Miller07},  developed for optical surveys, recently improved and used for the CFHTLensS~\cite{Miller13} and KiDS surveys.
As a proof of concept, we measure galaxies shapes and evaluate any shear bias in this new method, that we call \emph{RadioLensfit}~\cite{Rivi}, by simulating SKA1-MID  visibilities of individual galaxies at the phase centre.

\section{RadioLensfit Overview}
RadioLensfit is a chi-square fitting approach to measure radio galaxy shapes in the visibility domain. The galaxy model is defined only by the disc component as the analytical Fourier transform of a Sersic exponential brightness profile:
$I(r) = I_0 \exp(-r/\alpha)$,
linearly deformed according to ellipticity parameters \textbf{e} = $(e_1, e_2)$.
$I_0$ and $\alpha$ are respectively the central brightness and scalelength of the galaxy.
The likelihood is marginalised over non-interesting parameters such as flux, scalelength and position. This way we get a likelihood function of only the ellipticity that should be well described by a Gaussian.  
Galaxy shape measure is given by the mean likelihood and 1D standard deviation (defined as the square root of the covariance matrix determinant).  

The cosmic shear is estimated as a weighted average of the galaxies ellipticity. Weights take into account the variance of the ellipticity of the galaxy population and the 1D variance of each shape measure.
For a detailed description of the method see~\cite{Rivi}.
 
\section{Galaxy Distributions}
\label{priors}
The galaxy distributions used in our simulations are derived from the fitting to the 20~cm continuum radio observations taken with the VLA at a center frequency of 1400 MHz, covering a region of the Spitzer Wide-area InfraRed Extragalactic (SWIRE\footnote{http://heasarc.gsfc.nasa.gov/W3Browse/radio-catalog/vlasdf20cm.html}). 
For a detailed analysis of this online catalog see~\cite{OM2008}. 
We selected only faint sources with flux  $S \le 100\mu$Jy (more than 1500). 
 
\begin{figure}
\includegraphics[scale=0.35]{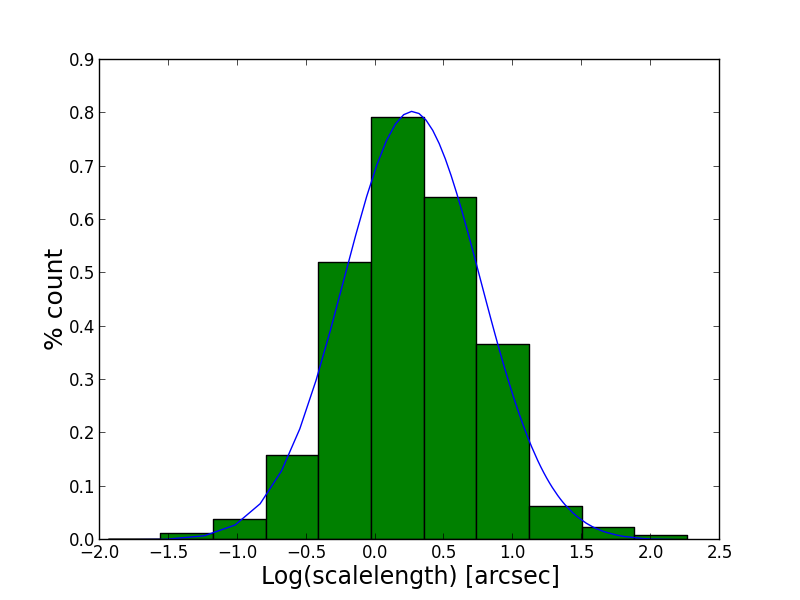}
\includegraphics[scale=0.35]{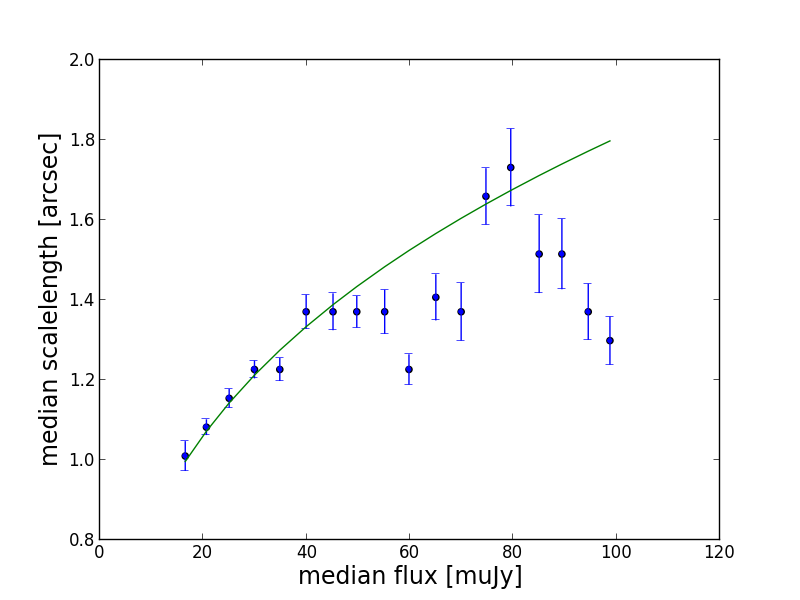}
\caption{Distributions for radio sources with flux below 100~$\mu$Jy in the SWIRE field observed by VLA. \emph{Left panel}: Observed scalelength histogram fitted by a lognormal distribution (solid curve). \emph{Right Panel}: Median scalelength versus median flux.}
\label{fig:scale}
\end{figure}

The resulting flux distribution is fitted by a power law: 
$p(S) \propto S^{-1.34}$, extrapolated below 30~$\mu$Jy because of incompleteness due to several detection effects (see \cite{OM2008}). 
The Sersic scalelength $\alpha$ of each galaxy is obtained from the relation FWHM =~$2\alpha\ln(2)$.
The left panel of Fig.~\ref{fig:scale} shows the scalelength distribution independently of the source flux, where data are well fitted by a lognormal distribution with mean 0.266 arcsec and standard deviation 0.496 arcsec. Note that this is consistent with the size distribution obtained from VLA+MERLIN observations of the HDF-North~\cite{Muxlow05}.
However, as claimed in \cite{Bondi08}, for a star-forming (SF) galaxy population there should be a linear relation between the log of the median major-axis scalelength $\alpha_{med}$ and flux density.  
As no source classification is provided in this catalog, we can estimate this relation only from sources with flux below 30~$\mu$Jy where such a relation appears (see right panel of Fig.~\ref{fig:scale}), as this population seems to be dominated by SF for this flux range. For higher fluxes there is fair mix of SF and (low-L) AGN as observed in the VLA-COSMOS survey~\cite{Sm2008} and therefore this relation is not evident anymore. For this reason we extrapolate the linear relation from a least-square fit to the median values in the range  $15\mu$Jy $\le S \le 30\mu$Jy, obtaining:
$\ln{[\alpha_{med}/\textrm{arcsec}]} = -0.93 +0.33\ln{[S/\mu \textrm{Jy}]},$
which is consistent with results in the literature.
As scalelength distribution dependent on the flux, we use a lognormal 
whose mean is $\mu=\ln(\alpha_{med})$ and variance is chosen in the middle of a range that appears to give a good representation of the distribution: $\sigma=0.3136$~arc sec.
As we have no information about ellipticity distributions on the radio regime, the modulus of the intrinsic ellipticity values are generated according to a distribution estimated from SDSS disk-dominated galaxies~\cite{Miller13}.

\section{SKA Simulations}
\label{sec:ska}
\begin{figure}
\includegraphics[scale=0.35]{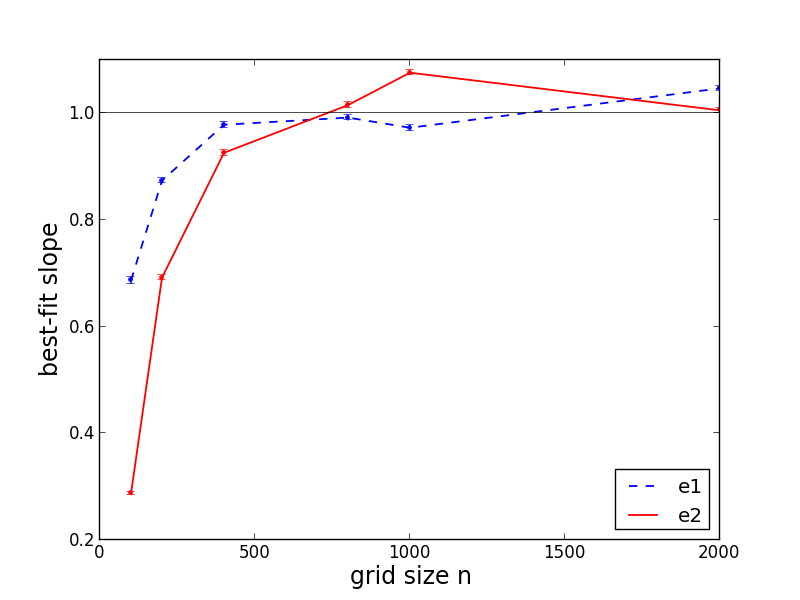}
\includegraphics[scale=0.35]{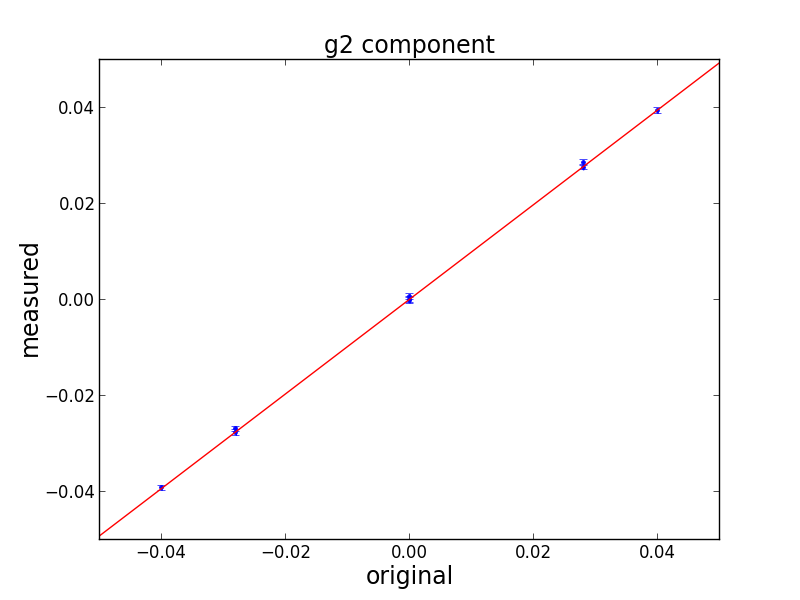}
\caption{\emph{Left panel}: Plot of the shape measurements best fitting slope vs grid size for a population of 1000 galaxies. \emph{Right panel}: Plot of the 2nd component of shear measurements and corresponding best-fit line.}
\label{fig:shear}
\end{figure}

By using the SKA1-MID baseline configuration, we simulated an 8-hour track radio observation at declination $\delta = 30^\circ$ of a population of individual SF galaxies. Visibilities are computed for the first 30\% of Band~2, i.e. 950 - 1190 MHz, and sampled every 60~s for 12 channels.   
Galaxies flux ranges in the interval 50~$\mu$Jy-100 $\mu$Jy corresponding to a SNR $\ge 40$. 
We apply a uniform gridding scheme to reduce the data volume and computational time. By testing the shape fitting, it appears that a small grid size $800 \times 800$ is the best choice for this case (see left panel of Fig.~\ref{fig:shear}).

To estimate the shear bias, we applied to each galaxy intrinsic ellipticity an input constant shear with amplitude $g=0$ and $g = 0.04$ (with 8 different orientations).
We compare input \textbf{g} and measured \textbf{g}$^m$ shear ellipticity values:
$g_i^m - g_i = m_i g_i + c_i$ ($i=1,2$),
where $m_i$ and $c_i$ are respectively the multiplicative and additive biases. They are
measured with an accuracy of 1\% by simulating for each measure a population of $10^4$ galaxies.
Shear bias estimates show different values for the two ellipticity components probably due to the asymmetry of the $uv$ coverage:
\begin{align}
& m_1 = -0.00079 \pm 0.00750,  \quad c_1 =  0.00014 \pm 0.00020; \nonumber \\ 
& m_2 = 0.0143 \pm 0.0073, \ \ \ \quad \quad c_2 =  0.00023 \pm 0.00019. \nonumber
\end{align} 

\section{Conclusions}
RadioLensfit is a method for shear measurement from radio weak lensing surveys working in the visibility domain.
We tested this method by simulating the visibilities of individual galaxies located at the phase centre using the SKA1-MID baseline configuration. RadioLensfit seems to be very promising for SKA continuum surveys because from these first tests we get multiplicative and additive noise biases comparable with the requirements on a 5000 $\deg^2$ SKA1 survey ($m = 0.0067$, $c = 0.00082$)~\cite{Brown15}. In particular the additive bias is on average 4 times smaller.
Further work will test this method for simulations with a lower SNR and with galaxies located randomly in the field of view taking into account frequency and time smearing effects.


\begin{thebibliography}{99}
\bibitem{Brown15}
M. Brown, D. Bacon, et al.
\emph{SKA Cosmology Chapter (AASKA14)}, arXiv:1501.03828, 2015.
\bibitem{Bondi08}
M. {Bondi}, P. {Ciliegi}, E. {Schinnerer, et al.}
\emph{ApJ}, 681:1129, 2008.
\bibitem{Miller07}
L.~{Miller}, T.~{Kitching}, C.~{Heymans}, et al.
\emph{MNRAS}, 382 (1): 315--324, 2007.
\bibitem{Miller13}
L.~{Miller}, C.~{Heymans}, T.~{Kitching}, et al.
\emph{MNRAS}, 429 (4): 2858--2880, 2013.
\bibitem{Muxlow05}
T.~{Muxlow}, A.~{Richards}, and S.~{Garrington, et al.}
\emph{MNRAS}, 358: 1159, 2005.
\bibitem{OM2008}
F.N.~{Owen}, G.E.~{Morrison}.
\emph{The Astronomical Journal}, 136: 1889, 2008.
\bibitem{Patel15}
P. Patel, I. Harrison, et al.
\emph {SKA Cosmology Chapter (AASKA14)}, arXiv:1501.03892, 2015.
\bibitem{Rivi}
M.~{Rivi}, L. {Miller}, S. {Makhathini}, F.B. {Abdalla},
\emph{in preparation}.
\bibitem{Sm2008}
V.~{Smolcic}, E. {Schinnerer}, M. {Scodeggio}, et al.
\emph{ApJS}, 177 (1): 14, 2008.

\end{thebibliography}
\end{document}